\documentclass[letterpaper,english,reprint, aps]{revtex4-1}
\usepackage[latin9]{inputenc}
\usepackage{amsmath}
\usepackage{amssymb}
\usepackage{graphicx}

\makeatletter

\newcommand*\LyXThinSpace{\,\hspace{0pt}}
\pdfpageheight\paperheight
\pdfpagewidth\paperwidth

\newcommand{\lyxmathsym}[1]{\ifmmode\begingroup\def\b@ld{bold}
  \text{\ifx\math@version\b@ld\bfseries\fi#1}\endgroup\else#1\fi}

\usepackage{cmbright}
\usepackage{lineno}
\makeatother

\usepackage{babel}
\begin{document}

\preprint{APS/123-QED}

\title{Transmission of heat modes across a potential barrier}

\author{Amir Rosenblatt$^{1,\dagger},$ Fabien Lafont$^{1,\dagger},$ Ivan
Levkivskyi$^{2},$ Ron Sabo$^{1},$ Itamar Gurman$^{1},$ Daniel Banitt$^{1},$
Moty Heiblum$^{1}$ and Vladimir Umansky$^{1}$}

\affiliation{$^{1}$Braun Center for Submicron Research, Dept. of Condensed Matter
physics, Weizmann Institute of Science, Rehovot 76100, Israel}

\affiliation{$^{2}$Institute of Ecology and Evolution, University of Bern, CH-3012
Bern, Switzerland}

\collaboration{$^{\dagger}$These authors contributed equally to this work}
\email{correspondence should be addressed to: lafont.fabien@gmail.com}

\maketitle

\section*{Abstract}
\textbf{Controlling the transmission of electrical current using a
quantum point contact constriction paved a way to a large variety
of experiments in mesoscopic physics. The increasing interest in heat
transfer in such systems fosters questions about possible manipulations
of quantum heat modes that do not carry net charge (neutral modes).
Here, we study the transmission of upstream neutral modes through
a quantum point contact in fractional hole-conjugate quantum Hall
states. Employing two different measurement techniques, we were able
to render the relative spatial distribution of these chargeless modes
with their charged counterparts. In these states, which were found
to harbor more than one downstream charge mode, the upstream neutral
modes are found to flow with the inner charge mode - as theoretically
predicted. These results unveil a universal upstream heat current
structure and open the path for more complex engineering of heat flows
and cooling mechanisms in quantum nano-electronic devices.}

\section*{Introduction}

The intimate link between heat current, entropy flow and therefore
information transfer \cite{Pendry1983,Sivan1986}, triggered recent
interest in thermoelectric effects occurring at the nanoscale, such
as measurements of the quantum limit of heat flow of a single quantum
mode \cite{Jezouin2013,Banerjee2017,Banerjee,Roukes2000}, heat Coulomb
blockade of a ballistic channel \cite{Sivre2017} or quantum limited
efficiency of heat engines and refrigerators \cite{Muhonen2012,Whitney2014}.
One main experimental obstacle in measuring thermal effects is to
decouple the charge from heat currents. Such separation is made easier
in the fractional quantum Hall effect (FQHE) \cite{Tsui1982}, since,
at least in hole-conjugate states (say, $1/2<\nu<1$), chargeless
heat modes propagate with an opposite chirality (upstream) to that
of the charge modes \cite{Bid2010,Altimiras,Venkatachalam2012}. Theoretically,
while Laughlin wave function \cite{Laughlin1983} has a great success
in describing various aspects of the FQHE at filling factors  $\nu=1/m$
, with odd $m$, the structure of hole-conjugate filling fractions,
such as $\nu=2/3$ and $\nu=3/5$, is still not clear. Two edge-model
structures had been proposed for the most studied $\nu=2/3$ state,
the first considers this state as a charge conjugate of the $\nu=1/3$
state; namely, a $\nu=1/3$ hole Landau level (LL) in the completely
filled $\nu=1$ electronic LL form a Laughlin condensate with $\nu=1/3$
\cite{MacDonald1990,Kane1994}. The second, considers it as a $\nu=1$
type condensate of fractional $e^{*}=1/3$ quasiparticles on top of
the $\nu=1/3$ state \cite{Meir1994,Wang2013}. Interestingly, recent
experiments with a softly defined edge potential (induced by a gate)
showed that the structure of the edge charge modes at filling factors
$\nu=2/3$ (and similarly at $\nu=3/5$) is in fact composed of two
spatially separated charge modes \cite{Sabo2017}, that would suggest
the second point of view to be more appropriate. Furthermore, topological
arguments require to have a conserved total number of net modes, therefore
predicting two upstream neutral modes at $\nu=2/3$, and three at
$\nu=3/5$.

Here, we fully characterize the transmission of the downstream charge
modes and the upstream neutral modes through a potential barrier,
imposed by a quantum point contact (QPC). We show their relative spatial
distribution and heat power carried by the neutral modes. Using the
same device, we shed more light on the interplay between charge and
neutral modes impinging on a QPC by distinguishing thermal fluctuations
from shot noise. 

\begin{figure*}
\centering{}\includegraphics[width=2\columnwidth]{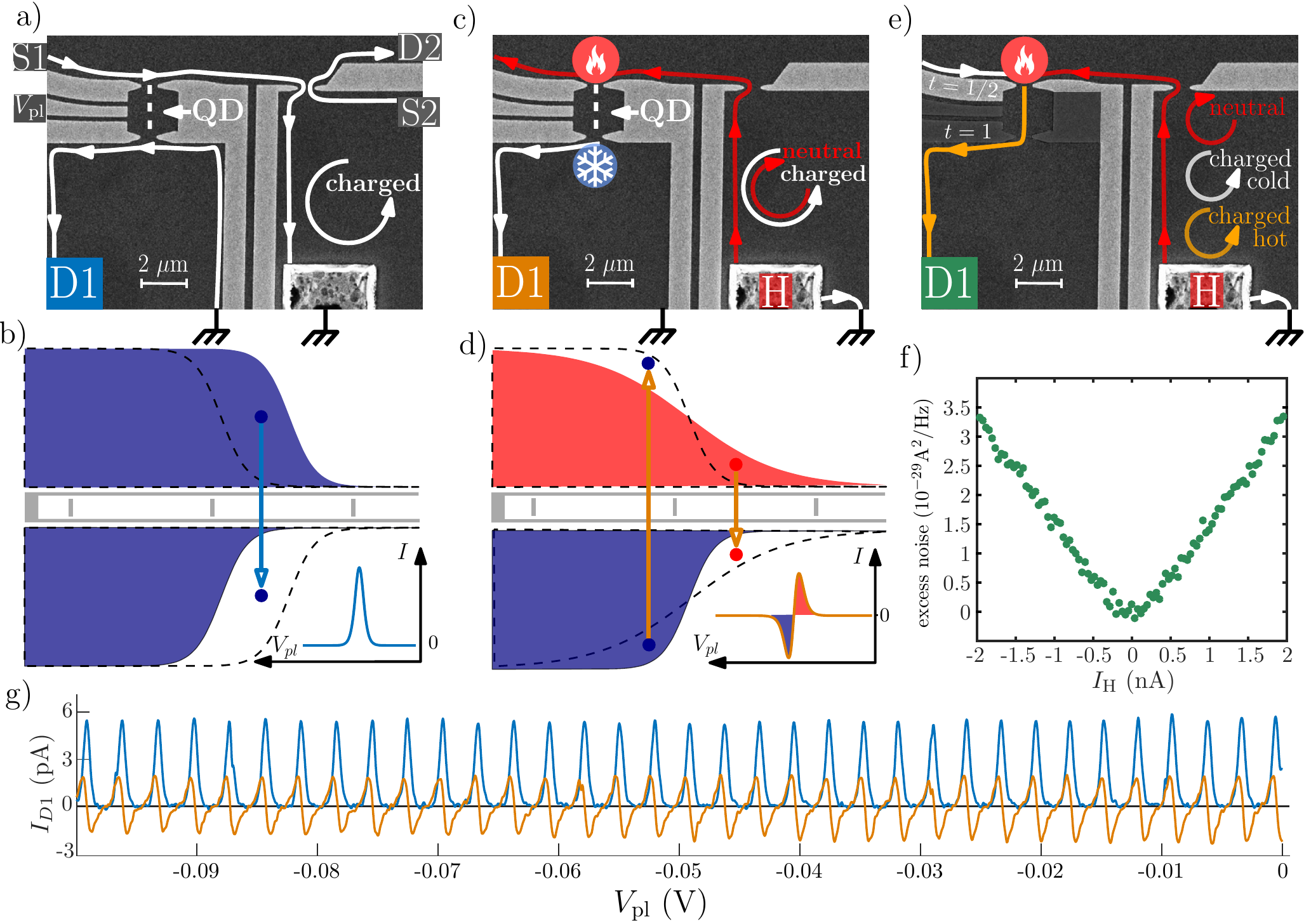}\caption{\textbf{Description of the experimental device and methods:} \textbf{a)}
Scanning electron microscope (SEM) image of the device. In this configuration
the current is sourced from S1, reaching to the quantum dot (QD).
Sweeping the plunger gate voltage results in a succession of Coulomb
peaks as shown by the blue curve on g). \textbf{b)} Schematic of the
equilibrium distribution on each side of the QD. When the plunger
gate is tuned, electron can tunnel into the dot from the high occupation
number region to the lower one, creating a positive current measured
at D1. \textbf{c)} Neutral mode heat detection configuration: The
current sourced in H is directed to the ground and plays no part in
the experiment. A hot spot present at the upstream side of contact
H excites the neutral heat modes flowing upstream towards the QD,
creating a thermal gradient across the QD. The produced thermoelectric
current then flows to D1. \textbf{d)} Sketch of the equilibrium distribution
on both side of the dot in the case depicted in c). In this case the
tunneling direction will depend if an energy level of the QD is placed
bellow or above the center of the distribution. This induces an alternating
current when the plunger gate is tuned. \textbf{e)} Noise measurement
configuration: Here the input QPC of the QD is set to half transmission
while the second one is fully open. This turns the QD into an effective
single QPC device. The heat carried by the neutral modes increase
the electron temperature at the input QPC of the dot, which increases
the Johnson-Nyquist noise, measured at D1. \textbf{f)} Excess noise
measured at D1 as function of current injected in H, as described
in e). \textbf{g)} Measurement corresponding to the configuration
a) in blue and c) in orange.}
\end{figure*}

\section*{Results}

\subsection*{Transmission of neutral modes accross a QPC}

The experimental setup, designed to map the transmission of neutral
modes through a QPC constriction, is presented in Figs. 1a, 1c, 1e.
The QPC (on the right) partitions either the charge mode emitted from
S2 (Fig. 1a), or the neutral modes excited by the hot spot at the
back of contact H (Figs. 1c \& 1e). In the latter case, two methods
were employed to convert the energy carried by the neutral modes to
a measurable charge current. The first utilized a quantum dot (QD,
on the left) to convert a temperature gradient to a net thermoelectric
current \cite{Beenakker1992,Staring1993,Viola2012,Gurman2012,Altimiras2010}
as sketched in Fig. 1c. With the thermal distribution at the input
of the QD being hotter than at the output, scanning the energy level
of the QD leads to a net thermoelectric current through the QD with
an alternating polarity (see Fig. 1d and Fig. 1g, and experimental
details in Methods Section). The second approach used excess noise
measurements, resulting from the upstream heat current impinging on
a QPC (Fig. 1e) \cite{Bid2010,Gross2012,Inoue2014}. Here, the input-QPC
of the QD was tuned to transmission half while the output-QPC was
fully open; therefore turning the QD to a single QPC. The impinging
neutral mode increases the electron temperature at the QPC and accordingly
the Johnson-Nyquist noise measured at D1 (Fig. 1f).

The thermopower of a QD, subjected to such a thermal gradient, was
studied both theoretically and experimentally \cite{Beenakker1990,Staring1993,Viola2012,Gurman2012}.
On the theoretical side, one needs to obtain the thermoelectric current
while considering a specific model of the edge modes structure (see
below). On the experimental side, we can directly relate the thermoelectric
current to the heat current carried by the transmitted neutral modes
(Fig. 2a). Since half of the injected power into contact H, $P_{\mathrm{H}}=1/2\times IV$,
is dissipated on the hot spot at the back of the ohmic contact H ,
it is proportional to the heat current (more details in sup. material),
$P\propto(T_{1}^{2}-T_{0}^{2})\times\pi^{2}k_{\mathrm{B}}^{2}/6h$,
carried away by the neutral modes which allows us to map the thermoelectric
current to the heat flow\cite{Jezouin2013,Banerjee2017}. Utilizing
this correspondence, the spatial distribution of the neutral modes
can be ascertained by measuring their transmission through a QPC constriction.
Applying constant power ($\sim$20 fW) to contact H and measuring
the thermoelectric (TE) current across the QD as we gradually pinch
the right-QPC (see details in Measurement technique), we find the
evolution of heat current carried by the neutral modes (\textquotedblleft neutral
transmission\textquotedblright ), that we present on Fig. 2c (blue
curve) together with the conductance of the QPC (black curve). The
neutral-transmission was extracted by taking the corresponding power
for a given measured TE current from the mapping in Fig.2a (blue curve);
normalized to a fully open right-QPC. A second measurement, using
the noise thermometry technique described above was used to validate
the neutral transmission. The results are presented on Fig. 2c-red
curve, where we plot the evolution of $T_{1}^{2}-T_{0}^{2}$ (which
is proportional to the heat current carried by the neutral modes,
see sup. material), where $T_{1}$ is measured and $T_{0}=30$ mK
is the base fridge temperature. It is clear from Fig.2c that the two
methods led to consistent results. Comparing the neutral transmission
with the one of the charge modes, one notices that $\sim$80\% of
the neutral mode power is reflected when the inner charge mode, with
conductance of $e^{2}/3h$, is reflected (at the beginning of the
conductance plateau, around $V_{\mathrm{QPC}}=-0.4$ V), showing that
most of the upstream heat flow is \textquoteleft attached\textquoteright{}
to the inner charge mode. A very similar result was obtained for $\nu=3/5$
with strong correlation between the reflection of the inner charge
mode and the reflection of the heat modes (Fig. 2e). 

\begin{figure*}
\centering{}\includegraphics[width=2\columnwidth]{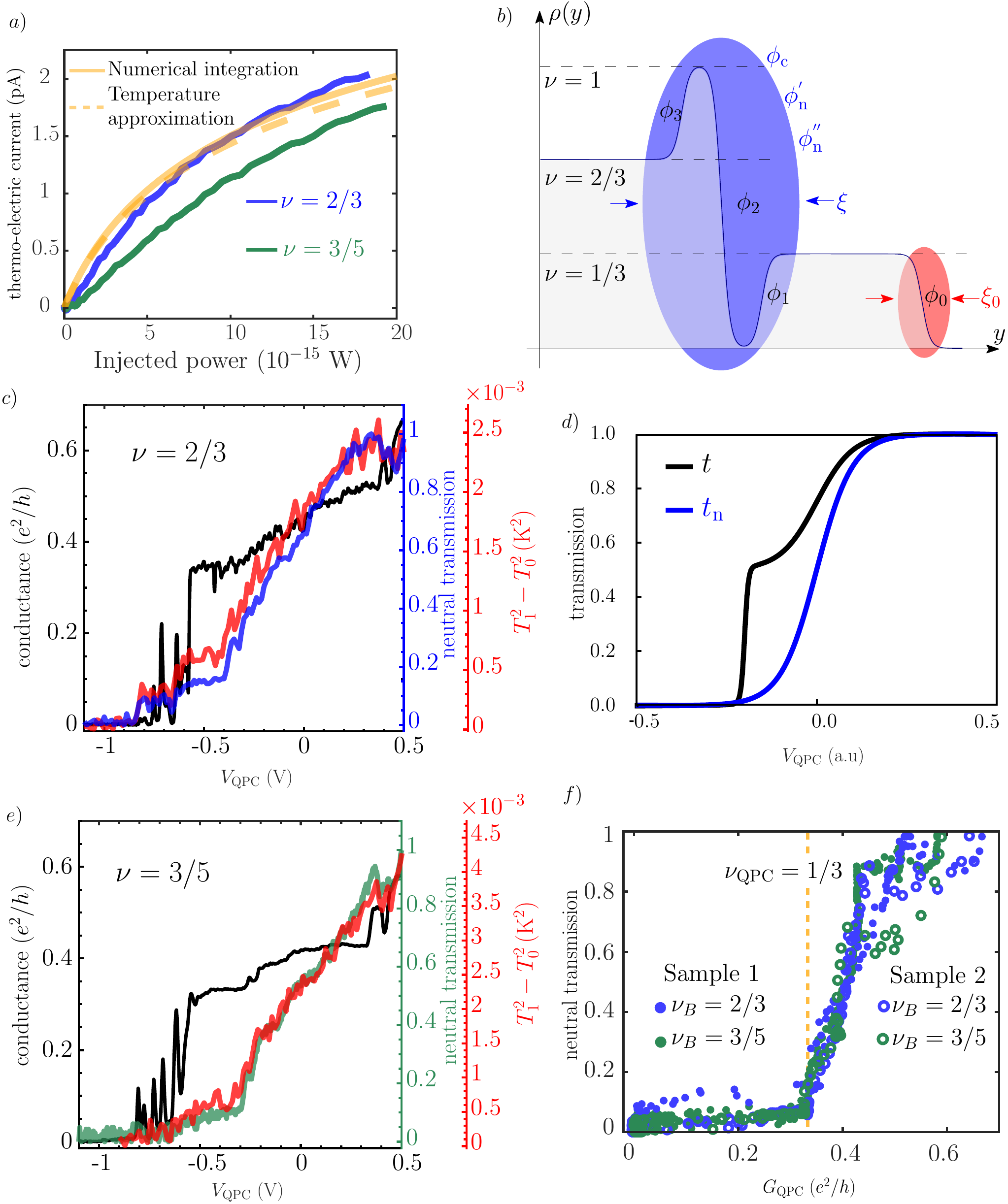}\caption{\textbf{Transmission of neutral modes: a)} Evolution of the thermoelectric
current as function of the Joule power applied on H at $\nu=2/3$
(blue curve) and at $\nu=3/5$ (green curve). Theoretical prediction
of the thermoelectric current through the QD, orange dashed curve
shows effective temperature approximation, orange solid curve shows
result of exact numeric integration. \textbf{b)} Theoretical model
of the edge structure of the $\nu=2/3$ state. Before equilibration
four charge modes $\phi_{0,1,2,3}$ are considered with respective
velocity set by $\partial\rho(y)/\partial y$. After renormalization
the system consists of two downstream charge modes of different width
$\xi$ and $\xi_{0}$ and two upstream neutral modes attached to the
inner mode. \textbf{c)} black curve-left axis: Conductance at the
QPC constriction as function of the QPC split gate voltage at $\nu=2/3$,
blue curve-right blue axis: Evolution of the neutral transmission
as function of the QPC split gate voltage. Red curve-red axis: evolution
of the excess noise measured at D1 as function of the QPC split gate
voltage. \textbf{d)} Theoretical charge and neutral transmission.
\textbf{e)} Similar to c) at filling factor $\nu=3/5$ . \textbf{f)}
Neutral transmission as function of the conductance in the QPC constriction
for the bulk filing factors $\nu_{\mathrm{B}}=2/3$ and $\nu_{\mathrm{B}}=3/5$.
The plain and open symbols designate two different samples.}
\end{figure*}

\subsection*{Theoretical Model}

We now compare our experimental results to a theoretical model of
the $\nu=2/3$ state. The first model for such state was developed
by MacDonald \cite{MacDonald1990}, that predicted two counter-propagating
charged edge modes with respective conductance $e^{2}/h$ flowing
downstream and $e^{2}/3h$ flowing upstream. Due to the absence of
any experimental evidence of upstream current \cite{Ashoori1992},
Kane, Fisher and Polchinski \cite{Kane1994,Kane1995} introduced scattering
between the above mentioned channels leading to a single downstream
charge mode with conductance $2e^{2}/3h$ and an upstream neutral
mode. Later, Meir \emph{et al.} \cite{Meir1994,Wang2013} refined
this model for a soft edge potential (which is the case for gate-defined
edge, like presently) and proposed an edge modes picture presented
on Fig. 2b. A smooth confining potential induces a non-trivial density
variation near the edge: Starting from the bulk, the local filling
factor goes from 2/3 to 1 creating an upstream charge mode of conductance
$e^{2}/3h$. The subsequent filling factor drop goes from unity to
0 and therefore creates a charge mode going downstream with an associated
conductance equal to $e^{2}/h$. Finally, an extra density hump reaching
$\nu=1/3$ creates a pair of $G=e^{2}/3h$ counter propagating channels.
Taking into account interactions and scattering between the channels
leads to the presence of a decoupled channel with conductance $e^{2}/3h$
of width $\xi_{0}$ close to the edge, flowing next to an inner, wider
channel of width $\xi$, also with conductance $e^{2}/3h$ accompanied
by two neutral modes flowing upstream. Note, that there are several
alternative pictures of QH states at filling factors 2/3 and 3/5 both
with and without neutral upstream edge modes (see, e.g., \cite{Wu2012}).
Moreover, edge reconstruction phenomena \cite{Chklovskii1992,Aleiner1994}
(e.g. additional humps in the density) can affect both the microscopic
and effective behavior of edge states. Therefore, we focus our discussion
on the simplest effective model of edge states that is less sensitive
to the details of experimental situation. The theoretical model presented
here utilizes the ``Meir \emph{et al.}'' edge structure where couplings
between channels leads to different excitation. The outer edge mode
is completely decoupled from all other channels, while the inner channel
gives rise to excitations, each characterized by a charge $e^{*}$
and a scaling dimension $\Delta$ (see details in Supplementary note
4), which allow us to calculate the TE current through a single energy
level $\epsilon_{0}$ of the QD  for this particular edge state picture
and for dominant excitation with $e^{*}=1/3$ and $\Delta=1$\cite{Levkivskyi2009}:\small
\begin{equation}
I_{\mathrm{TE}}\propto[f_{\mathrm{out}}(T_{\mathrm{out}},\epsilon_{0})f_{\mathrm{in}}(T_{\mathrm{in}},-\epsilon_{0})-f_{\mathrm{in}}(T_{\mathrm{in}},\epsilon_{0})f_{\mathrm{out}}(T_{\mathrm{out}},-\epsilon_{0})]
\end{equation}
\normalsize

where $f_{\mathrm{out}},\,f_{\mathrm{in}},T_{\mathrm{out}},T_{\mathrm{in}}$
are the occupation numbers and temperatures corresponding to the Input
and Output side of the quantum dot. Considering that the two upstream
modes originate from a hot reservoir at temperature $T_{1}$ and the
downstream ones at temperature $T_{0}$, it is possible to express
the QD's input/output occupation number as, 

\scriptsize

\begin{equation}
f_{\mathrm{in}}(\epsilon_{0})=\int dt\,e^{i\epsilon_{0}t}\left[\frac{T_{0}}{\sinh(\pi T_{0}(t-i\eta))}\right]^{\delta_{0}}\left[\frac{T_{1}}{\sinh(\pi T_{1}(t-i\eta))}\right]^{\delta_{1}}
\end{equation}

\normalsize

with $\ensuremath{\delta_{0}=1/3,\delta_{1}=2/3}$ , $\ensuremath{\Delta=\delta_{0}+\delta_{1}=1}$
and 

\small
\begin{equation}
f_{\mathrm{out}}(\epsilon_{0})=\int dt\,e^{i\epsilon_{0}t}\frac{T_{0}}{\sinh(\pi T_{0}(t-i\eta))}=\frac{1}{e^{\epsilon_{0}/T_{0}}+1}
\end{equation}

\normalsize

\begin{figure*}
\centering{}\includegraphics[width=2\columnwidth]{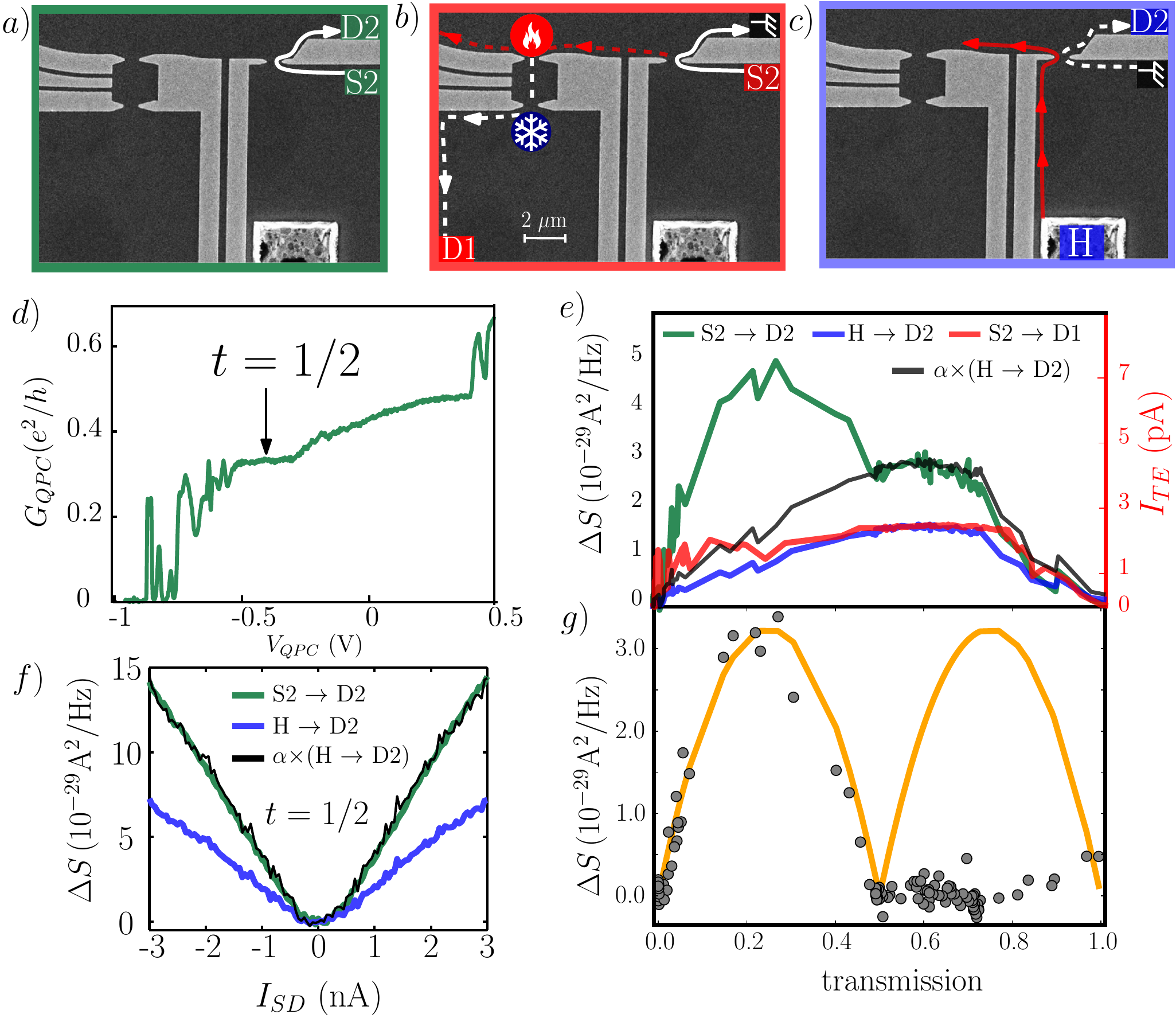}\caption{\textbf{Noise measurements: a)} Sketch of noise measurement at D2
when sourcing charge current from S2 \textbf{b)} Sketch of thermoelectric
measurement at D1 when sourcing from S2, the impinging current generates
heat carried upstream by the neutral modes and converted to thermoelectric
current at the QD. \textbf{c)} Sketch of noise measurement at D2 when
sourcing from H, the upstream neutral modes excited at the hot spot
increase the Johnson\textendash Nyquist noise measured at D2 \textbf{d)}
Conductance of the QPC as function of the split gate voltage.\textbf{
e)} Green: Current fluctuations measured at D2 when sourcing from
S2 as function of the transmission of the QPC. Blue: Current fluctuations
measured at D2 when sourcing from H. black: Same as blue multiplied
by a scaling factor $\alpha=2$. Red: Thermoelectric current measured
at D1 when sourcing from S2. \textbf{f)} Excess noise measured at
D2 as function of the current at transmission half when sourcing from
S2 (green) or H (blue). The black curve represents the noise from
H to D2 (blue curve) multiplied by the scaling factor $\alpha=2$.\textbf{
g)} Grey dots: Shot noise contribution $N_{\mathrm{SN}}=N_{\mathrm{tot}}-\alpha N_{\mathrm{th}}=N_{\mathrm{S2\rightarrow D2}}-\alpha N_{\mathrm{H\rightarrow D2}}$.
Orange curve: expected shot noise from 2 separated ballistic charge
channels.}
\end{figure*}

Inserting Eqs. (2) \& (3) in Eq. (1) one can calculate the evolution
of the TE current in a QD with a single energy level for this particular
edge modes picture (see Supplementary note 4). The results either
using effective temperature approximation or exact numeric integration
are plotted in Fig.2a. The good agreement between the theoretical
and experimental results strengthen the validity of an edge picture
that consists of two charge modes accompanied by two upstream neutral
modes. Using the same theoretical picture, we have modeled the neutral
and charge transmissions trough the QPC constriction using a quasi-classical
approximation \cite{Buttiker1990}: 
\begin{equation}
t_{i}(V)=1/\left(1+e^{(V-V_{i})/\delta V_{i}}\right)
\end{equation}
where $i=$inner, outer. In our model there are two modes that are
spatially separated, therefore they will be centered around different
gate voltage $V_{i}$ of the split-gate QPC, and with different width
corresponding to $\delta V_{i}$. The total charge transmission $t_{\mathrm{charge}}(V)=\left(t_{\mathrm{inner}}(V)+t_{\mathrm{outer}}(V)\right)/2$
is plotted in Fig.2d together with the neutral transmission, $t_{\mathrm{neutral}}(V)=t_{\mathrm{inner}}(V)$,
where both the neutral modes are located near the inner charge mode.
As visible on Fig.2d, this simple model is able to qualitatively explain
the transmission of the neutral modes; and in particular the observed
vanishing of the neutral transmission when the conductance reaches
$G_{\mathrm{QPC}}=1/3$. To compare in more details the measured transmission
of the neutral modes in the two bulk fractions, $\nu_{\mathrm{B}}=2/3$
and $\nu_{\mathrm{B}}=3/5$, we have plotted each one as function
of the conductance of the QPC constriction $G_{\mathrm{QPC}}$ (Fig.2f).
Surprisingly, they present a nearly identical behavior, which strongly
suggests that, at least for these two hole-conjugate states, the transmission
of the neutral modes is dictated by the conductance of the QPC and
is poorly dependent on the bulk filling factor. We have plotted on
the same figure (open symbols) the results of another similar sample
(presented in Supplementary note 1). It is clear that the curves present
an overlap on a large $G_{\mathrm{QPC}}$ region which indicates that
the structure of the neutral modes is universal and do not depend
on the bulk state as well as the particular mobility or disorder in
the QPC constriction. 

\subsection*{Excess noise subtraction}

Noise measurements reveal even more about energy exchange mechanisms
between the charge and neutral modes. We first start from the configuration
presented on Fig.3a, where we measure excess-noise in D2 when sourcing
DC current from S2. As shown on Figs. 3e \& 3f (green curves) and
reported previously \cite{Saminadayar1997,Bid2009,Gross2012,Inoue2014},
the noise remain finite when the QPC is tuned to the conductance plateau
at $t_{\mathrm{QPC}}=1/2$ with $G_{\mathrm{QPC}}=e^{2}/3h$ (Fig.
3d). This finite noise is even more puzzling since in recent experiment
\cite{Sabo2017} it was shown that two spatially separated channels
are present, excluding the presence of conventional shot noise at
such transmission. Taking advantage of this device we were able to
show that a thermal contribution is added to the shot noise. In order
to distinguish between the thermal noise and the partition noise of
the quasiparticles (shot noise), the TE current in the QD was measured
as function of the right-QPC, but this time the current was sourced
in S2 (Fig. 3b). The finite measured thermoelectric current presented
on Fig. 3e (red curve) is a clear signature of heat dissipation occurring
at the right-QPC constriction where the charge modes exchange energy
with the neutral mode. Similarly, exciting neutral modes using contact
H as shown in Fig. 3c, increases locally the electronic temperature
at the QPC \textendash{} as discussed before (Fig. 3f, blue curve).
Indeed, a similar dependence on the right-QPC transmission, in both
measurements, is observed in Fig. 3e. Subsequently, it is interesting
to exclude the thermal contribution from the total excess noise measured
at D2 when sourcing from S2 and be left with the shot-noise contribution
solely. In order to do that, we subtracted the thermal noise $N_{th}$
multiplied by a constant $\alpha$, that characterizes the neutral
mode decay, from the total noise $N_{\mathrm{tot}}$. The constant
$\alpha$ was chosen such that the shot noise contribution at half-transmission
would be zero - as expected from a full transmission of one channel
and a full reflection of the second. The result of this subtraction
$N_{\mathrm{sn}}=N_{\mathrm{tot}}-\alpha N_{\mathrm{th}}$ is plotted
on Fig. 3g together with the expected excess noise (orange curve)
for two charged channels giving a \textquotedblleft double hump\textquotedblright{}
shape (due to the $t(1-t)$ dependence of the noise of each channel).
The shot noise $N_{\mathrm{SN}}$ follows the expected behavior in
the range $0-0.5$ of total QPC transmission, but then collapses to
zero in all the region above half transmission. This would imply that
the measured noise when the inner mode is being partitioned results
solely from thermal fluctuations - consistent with the neutral modes
being attached to the inner channel. This would furthermore indicate
that, the outer mode can be partitioned like a ballistic channel while
the inner one exhibits dissipation. More experimental and theoretical
studies of this excess noise are required to have a full picture of
such state. 

\section*{Discussion}

Via studying the transmission of neutral modes through a QPC constriction
at hole conjugate states, $\nu=2/3$ and $\nu=3/5$ we showed that
this structure is consistent with the presence of two upstream neutral
modes attached to the downstream inner charge mode; being in agreement
with the proposed model by Meir \cite{Meir1994,Wang2013}. Moreover,
this transmission of the neutral modes through a barrier formed by
a QPC constriction is governed by its conductance and does not depend
on the bulk-filling factor. This suggests a universality of the neutral
modes morphology, which should guide future theoretical models describing
them and their interplay with the charge modes. More generally, these
results mark a new step in the understanding of heat transport in
the FQHE regime as well as for possible future implementation of controlled
engineering of heat currents on nanoscale electronic devices.

\section*{Methods}

\subsubsection*{Sample fabrication}

The samples were fabricated in GaAs\textendash AlGaAs heterostructures,
embedding a 2DEG, with an areal density of $(1.2\lyxmathsym{\textendash}2.5)\times10^{11}\LyXThinSpace\mathrm{cm^{-2}}$
and a 4.2\LyXThinSpace K ``dark'' mobility$(3.9\lyxmathsym{\textendash}5.1)\times10^{6}\mathrm{cm^{2}V^{-1}s^{-1}}$,
70\textendash 116 nm below the surface. The different gates were defined
with electron beam lithography, followed by deposition of Ti/Au. Ohmic
contacts were made from annealed Au/Ge/Ni. The sample was cooled to
30\LyXThinSpace mK in a dilution refrigerator.

\subsubsection*{Measurement technique}

Conductance measurements were done by applying an a.c. signal with
$\sim1\mu$V r.m.s. excitation around 1.3 MHz in the relevant source.
The drain voltage was filtered using an LC resonant circuit and amplified
by homemade voltage preamplifier (cooled to 1\LyXThinSpace K) followed
by a room-temperature amplifier (NF SA-220F5).

For the thermoelectric current measurement, an alternating voltage
$V_{\mathrm{H}}$ at frequency f$\approx$650 kHz was applied to contact
H, giving rise to an upstream neutral modes  with temperature proportional
to $|V_{\mathrm{H}}|$, producing a series of harmonics starting at
$2f$. The heat reaching the upper side of the QD generates an alternating
thermoelectric current flowing to drain D1. The signal is then filtered
using an LC circuit with a resonant frequency at $2f=1.3$ MHz amplified
by a home-made voltage cold amplifier followed by a room-temperature
amplifier, and finally reaching a lock-in amplifier set to measure
at frequency 2f. The thermoelectric current was measured by sweeping
the plunger gate for a given $V_{\mathrm{H}}$ amplitude by averaging
the peak-to-peak voltage of more then 20 consecutive oscillations.
The QD is in the metallic regime $\Delta E\ll k_{\mathrm{B}}T\ll e^{2}/C$
where the temperature is above the level spacing $\Delta E$ and below
the charging energy $e^{2}/k_{\mathrm{B}}C=290\text{ mK}$ (see Coulomb
diamond measurement in Supplementary note 3).

\subsection*{Data availability}

The data that support the plots within this paper and other findings
of this study are available from the corresponding author upon request. 

\subsection*{Acknowledgments}

A.R and F.L acknowledge Robert Whitney, Ady Stern, Yuval Gefen, Jinhong
Park and Kyrylo Snizhko for fruitful discussions. M.H. acknowledges
the partial support of the Minerva Foundation, grant no. 711752, the
German Israeli Foundation (GIF), grant no. I-1241- 303.10/2014, the
Israeli Science Foundation, grant no. ISF- 459/16 , and the European
Research Council under the European Community\textquoteright s Seventh
Framework Program (FP7/2007-2013)/ERC Grant agreement No. 339070. 

\subsection*{Author contributions}

A.R., F.L., R.S., I.G., and M.H. designed the experiment, A.R., F.L.,
R.S., I.G., D.B. preformed the measurements, A.R., F.L., R.S., I.G.,
D.B. and M.H. did the analysis, I.L. developed the theoretical model.
A.R., F.L., I.L. and M.H. wrote the paper. V.U. grew the 2DEG.

\subsection*{Competing Financial Interest}

The authors declare no competing financial interests.

\bibliographystyle{naturemag_NoUrl}

\end{document}

\section*{Supplementary Note 1: Eliminating bulk thermal transport}

Here we present a second sample studied in order to probe if any contribution
to the heat measured by the quantum dot was propagating through the
bulk and not only through the edge modes. The device is very similar
to the one from the main text. A deflector gate is placed before the
quantum dot that allows to redirect the neutral modes to the ground
directly. The results of the thermoelectric current measured versus
the plunger gate voltage for the two configurations, deflector open
and close, are presented on Supplementary Figure 1. When the QPC is closed a net thermoelectric
current is measured similarly to the main paper sample. Nevertheless,
one can notice that the thermoelectric current is only positive, which
is due to the fact that we used a spectrum analyzer, only sensitive
to the absolute value of the signal. Conversely, when the deflector
is open, no thermoelectric current is measured by the quantum dot,
which shows that no measurable contribution arise from the bulk.

\begin{figure}[H]
\begin{centering}
\includegraphics[width=0.72\columnwidth]{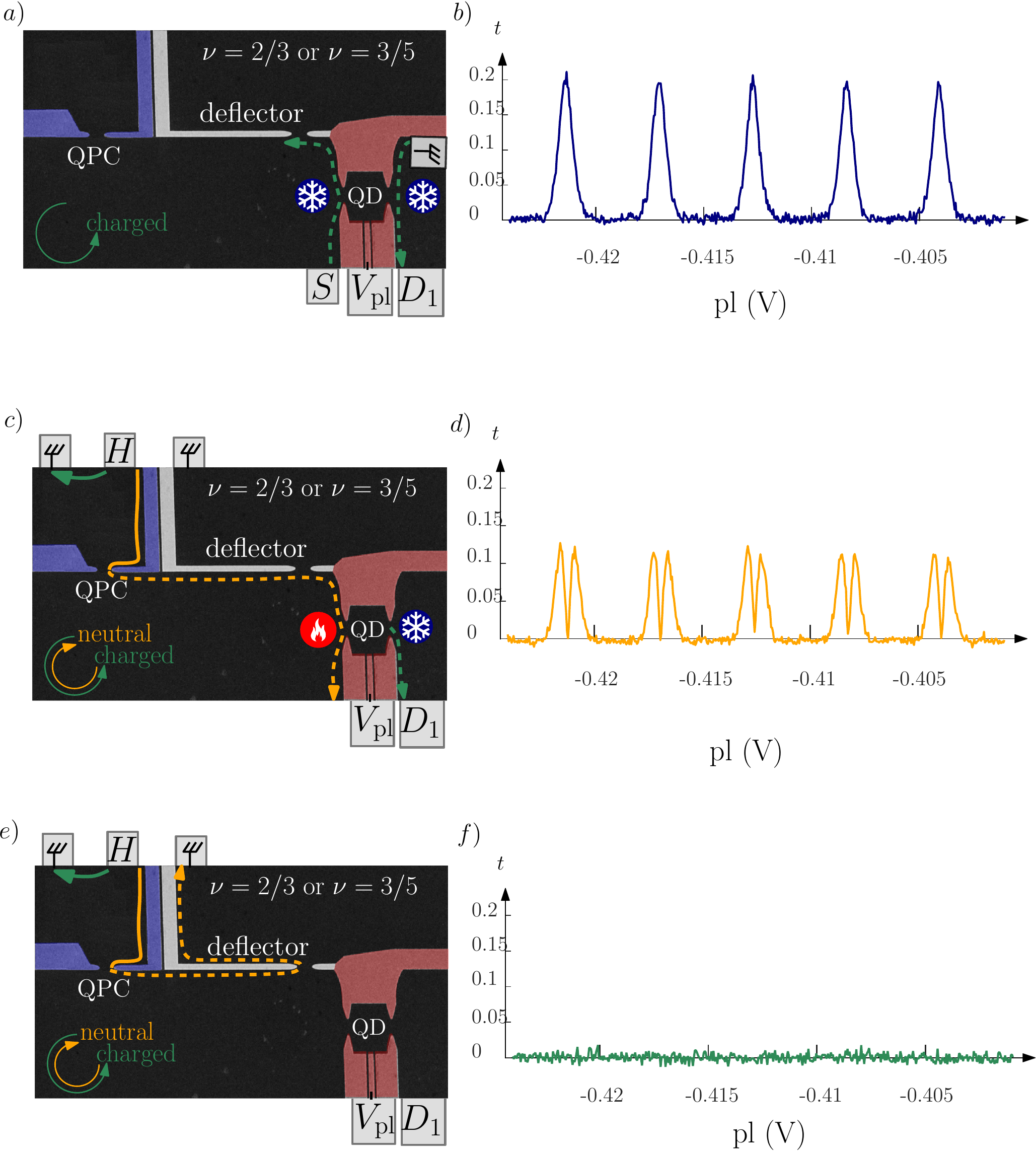}
\par\end{centering}
\caption{\textbf{Measurement of the second sample: a)} Configuration corresponding
to the measurement of the Coulomb blockade peaks. both sides of the
quantum dot are at base temperature. Sourcing current from S results
in transmission peaks measured in D1 when changing the plunger gate
voltage, as visible on \textbf{b)}. \textbf{c)} Configuration corresponding
to the measurement of the neutral transmission using the thermoelectric
current created through the quantum dot when the deflector gate is
closed. \textbf{d)} Evolution of the thermoelectric current across
the quantum dot as function of the plunger gate voltage. \textbf{e)}
Configuration corresponding to the measurement of the neutral transmission
using the thermoelectric current created through the quantum dot when
the deflector gate is open\textbf{ f)} No thermoelectric current is
measurable through the QD when the deflector is open showing that
no heat transport is happening in the bulk of the sample.}
\end{figure}

\newpage{}

\section*{Supplementary Note 2: Upstream neutral modes at other filling factors}

In Supplementary Figures 2 we present thermoelectric measurements at other filling
factors, $\nu=2,\,1,\,2/3,\,3/5,\,2/5,\,1/3$ together with the coulomb
blockade (CB) conductance as function of the plunger gate. At all
of the measurements we sourced AC voltage of $V_{\mathrm{H}}\sim70\,\mu\text{V}_{\text{RMS}}$
and scanned the plunger over a range of CB peaks where the QPC if
fully open and the deflector gate is energized. At fillings $\nu=2,\,2/5$
and $1/3$ there were no measurable thermoelectric voltage, consisting
with the lack of upstream neutral modes at these states. At fillings
$\nu=1,\,2/3$ and $3/5$ we measure significant thermoelectric voltage
consisting with the upstream neutral expected in the hole-conjugate
states ($\nu=2/3$ and $3/5$). At filling $\nu=1$ upstream heat
mode was measured before \citep{Venkatachalam2012,Gurman2016} and
it was attribute to a state of $\nu=2/3$ underlying the $\nu=1$.

\begin{figure}[H]
\begin{centering}
\includegraphics[width=1\textwidth]{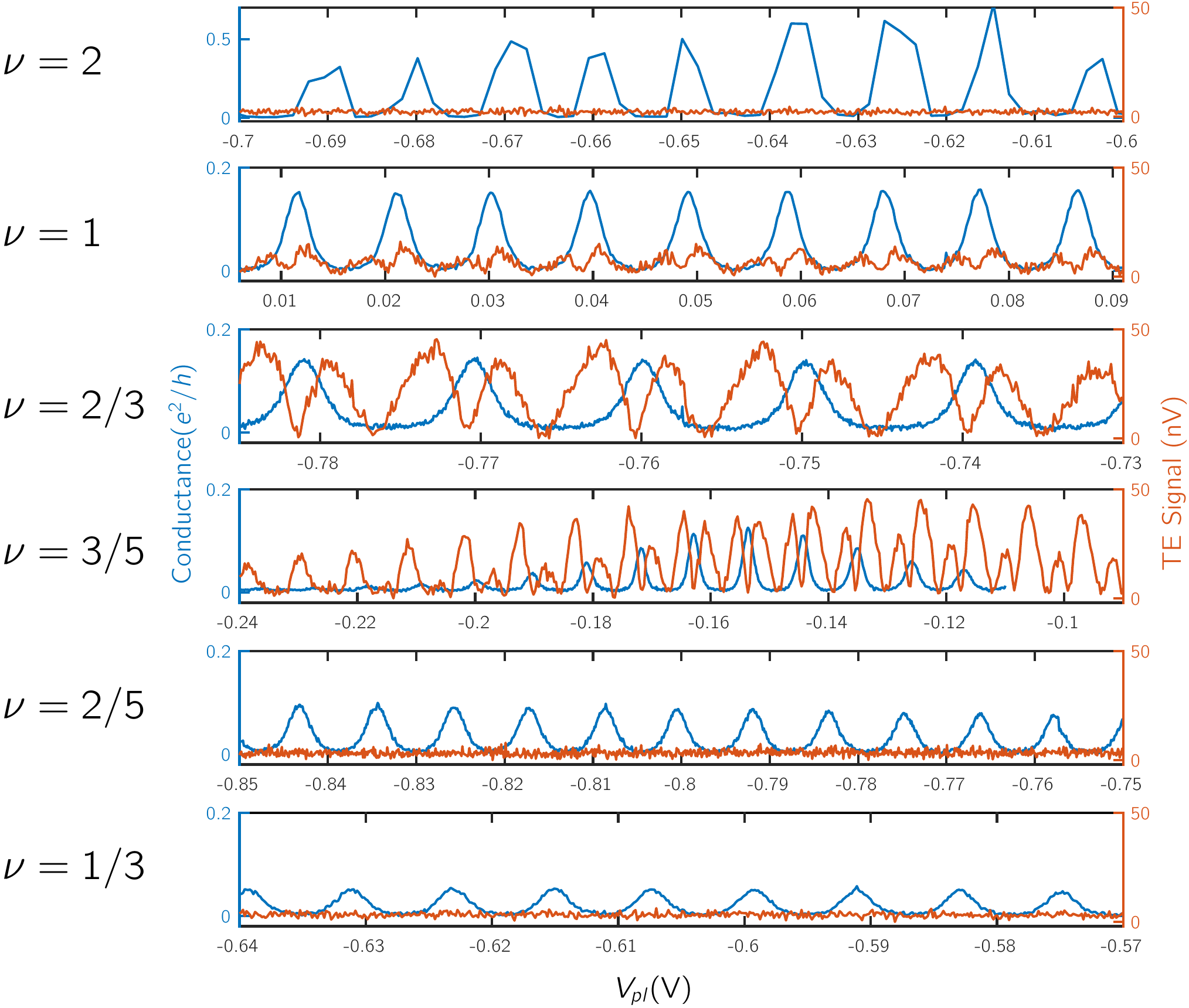}
\par\end{centering}
\caption{\textbf{Evolution of the Coulomb peaks and thermoelectric currents
for several filling factors:} Thermoelectric voltage at other filling
factors, $\nu=2,\,1,\,2/3,\,3/5,\,2/5,\,1/3$. Blue curves - left
axes - conductance through the QD tuned to Coulomb blockade regime.
Orange curves - right axes - thermoelectric voltage at the same regime.
Measured with the deflector gate energized and $V_{\mathrm{H}}\sim70\,\mu\text{V}_{\text{RMS}}$}
\end{figure}

\newpage{}

\section*{Supplementary Note 3: Quantum Dot Coulomb diamonds}

Here is presented the evolution of the conductance in the quantum
dot as function of the voltage applied in S1 and the plunger gate.
The extracted charging energy is $\sim25\,\mu e$V which corresponds
to an equivalent temperature of $\sim290$ mK.

\begin{figure}[H]
\begin{centering}
\includegraphics[width=1\textwidth]{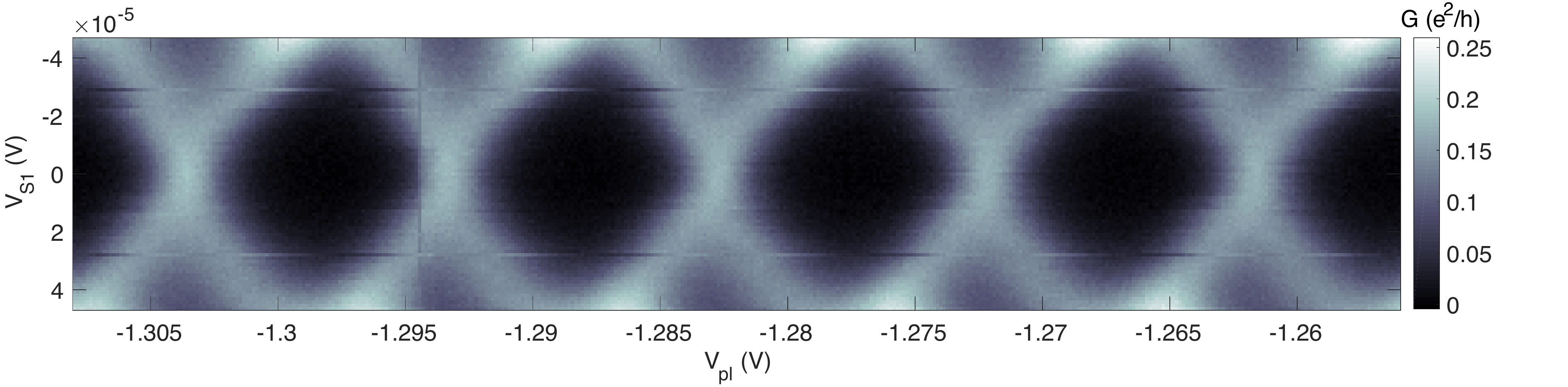}
\par\end{centering}
\caption{\textbf{Coulomb Diamonds:} Evolution of the conductance of the quantum
dot as function of $V_{\mathrm{S1}}$ and $V_{\mathrm{pl}}$. Coulomb
diamonds are visible with a typical charging energy of $25\,\mu e$V
with corresponds to an equivalent temperature of $\sim290$ mK}
 
\end{figure}

\section*{Supplementary Note 4:Theoretical model }

\label{model-sec} Fractional quantum Hall states at simple Landau
level filling fractions like $\nu=1/m$ with odd $m$ could be described
by the Laughlin wave function \citep{Laughlin1983}: 
\begin{equation}
\langle z_{1},...,z_{n}|\Psi_{\mathrm{L}}\rangle\propto\prod_{i<j}(z_{i}-z_{j})^{m}\exp[-\frac{1}{4l_{\mathrm{B}}^{2}}\sum_{i}|z_{i}|^{2}],\label{laugh}
\end{equation}
where $z_{i}=x_{i}+iy_{i}$ are complex coordinates of electrons in
the plane of two-dimensional electron gas and $l_{\mathrm{B}}=\sqrt{\hbar c/eB}$
is the magnetic length. This state features fractionally charged anyons
that are created by the operator $\psi^{\dag}(\zeta)$ defined via
\begin{equation}
\langle z_{1},...,z_{n}|\psi(\zeta)|\Psi_{\mathrm{L}}\rangle=\prod_{i}(\zeta-z_{i})\langle z_{1},...,z_{n}|\Psi_{\mathrm{L}}\rangle.\label{exc1m}
\end{equation}
The Laughlin wave function (\ref{laugh}) has a great success in describing
various aspects of the corresponding fractional quantum Hall states.
It has also been shown to be very close to exact eigenstate by numeric
calculations and shown to be a universal exact wave function for short
range interactions.

For more complex filling fractions like $\nu=2/3,3/5$, etc., it is
still not completely clear what is the structure the wave function.
Here we focus on the $\nu=2/3$ state and discuss consequences for
other filling factors later. There are two points of view on the $\nu=2/3$
state. One considers this state as a charge conjugate to the $\nu=1/3$
state. In other words, holes in the completely filled Landau level
$\nu=1$ form a Laughlin condensate with $\nu=1/3$. Another one considers
$\nu=2/3$ state as a $\nu=1$ type condensate of fractional $e^{*}=1/3$
quasi-particles on top of the $\nu=1/3$ state. It can be described
by the wave function: 
\begin{equation}
|\Psi_{\frac{2}{3}}\rangle\sim\int J(\zeta,\bar{\zeta})d^{M}\zeta d^{M}\bar{\zeta}\prod_{i<j}(\zeta_{i}-\zeta_{j})\exp[-\frac{1}{4ml_{\mathrm{B}}^{2}}\sum_{i}|\zeta_{i}|^{2}]\prod_{i}\psi^{\dag}(\zeta_{j})|\Psi_{\mathrm{L}}\rangle.\label{s23}
\end{equation}
Note that the magnetic length of inner condensate is renormalized
by $\sqrt{m}$. Such construction in general could describe various
filling factors in the hierarchic manner \citep{Prange}.

We argue that the present experiment indicates that the second point
of view is more appropriate. In order to investigate the edge structure
of such state (\ref{s23}), we take into account that the energy scales
are much lower than the fractional gap and use the effective low-energy
theory \citep{Halperin1982,Wen1990,FROHLICH1991517} to describe the
edge states. It has been shown \citep{Levkivskyi2016} that the generic
action for Abelian effective model of edge sates can be always cast
in the form: 
\begin{equation}
S[\phi_{s}]=\frac{1}{4\pi}\sum_{s}\int dtdx[\sigma_{s}D_{t}\phi_{s}D_{x}\phi_{s}-v_{s}(D_{x}\phi_{s})^{2}+Q_{s}\epsilon^{\mu\nu}a_{\mu}\partial_{\nu}\phi_{s}],\label{action}
\end{equation}
where $\mu=x,t$, $\sigma_{s}=\pm1$ denotes chirality of the corresponding
eigenmode, and covariant derivative $D_{\mu}\phi_{s}=\partial_{\mu}\phi_{s}+\sigma_{s}Q_{s}a_{\mu}$
depends on the couplings $Q_{s}$ of the corresponding modes to the
external electro-magnetic potential $a_{\mu}$. The charge density
operator in such theory is expressed in terms of boson fields as $\rho_{s}=(\sigma_{s}Q_{s}/2\pi)\partial_{x}\phi_{s}$.
There has been proposed an edge reconstruction picture \citep{Wang}
as shown in Fig S3. Here we follow this idea and propose a particular
model with $\hat{Q}=(1/\sqrt{3},1/\sqrt{3},0,0)^{T}$, $\sum_{s}\sigma_{s}Q_{s}^{2}=\nu$
and electron fields: \begin{subequations} 
\begin{eqnarray}
\phi_{1} & = & \sqrt{3}\phi_{{\rm c}}+\sqrt{\frac{3}{2}}\phi'_{{\rm n}}+\frac{1}{\sqrt{2}}\phi''_{{\rm n}}\\
\phi_{2} & = & \sqrt{3}\phi_{{\rm c}}+\sqrt{2}\phi''_{{\rm n}}\\
\phi_{3} & = & \sqrt{3}\phi_{{\rm c}}-\sqrt{\frac{3}{2}}\phi'_{{\rm n}}+\frac{1}{\sqrt{2}}\phi''_{{\rm n}}
\end{eqnarray}
\label{els} \end{subequations} 

Note that such model differs from proposed in Ref.\ \citep{Wang}
only in the degenerate subspace of the neutral modes. In the limit
of strong interactions the velocity of charged mode is large $v_{{\rm c}}\gg v'_{{\rm n}},v''_{{\rm n}}$.
We speculate that the width of inner reconstructed modes is also large
$\xi/\xi_{0}\gg1$. This is consistent with the fact that the correlation
length in the inner condensate of (\ref{s23}) is larger. Statistical
phases for electronic excitation are indeed fermionic $\theta_{\alpha\beta}=\pi K_{\alpha\beta}$,
$\alpha,\beta=0,...,3$. 
\begin{equation}
\hat{K}=\left(\begin{array}{cccc}
3 & 0 & 0 & 0\\
0 & 1 & 2 & 4\\
0 & 2 & 1 & 2\\
0 & 4 & 2 & 1
\end{array}\right)
\end{equation}
so that electron vectors form an integral lattice that has a hexagonal
projection on the neutral sector. Quasi-particles must have single-valued
wave functions as, e.g., Laughlin quasi-particles (\ref{exc1m}) discussed
above. In the effective model language this translates into integer
statistical phases with respect to all electronic excitation. In other
words, quasi-particles form a dual lattice: 
\begin{equation}
\chi_{\mathbf{n}}=\frac{n_{0}}{\sqrt{3}}\phi_{0}+\frac{n_{1}-n_{2}+n_{3}}{\sqrt{3}}\phi_{{\rm c}}+\frac{n_{3}-n_{1}}{\sqrt{6}}\phi'_{{\rm n}}+\frac{n_{1}-2n_{2}+n_{3}}{\sqrt{2}}\phi''_{{\rm n}}
\end{equation}
The charges of the corresponding excitation are $e_{\mathbf{n}}^{*}=(n_{0}+n_{1}-n_{2}+n_{3})/3$
and the scaling dimensions are given by: 
\begin{equation}
\Delta_{\mathbf{n}}=\frac{1}{3}n_{0}^{3}+n_{1}^{2}-\frac{1}{3}n_{2}^{2}+n_{3}^{2}+\frac{4}{3}n_{1}n_{3}-\frac{8}{3}n_{2}(n_{1}-n_{2}+n_{3}).
\end{equation}

It is interesting to investigate the most relevant particles content
in the inner channel, we index them as $\chi_{n_{1}n_{2}n_{3}}$.
There is one excitation with $e^{*}=1/3$ and $\Delta=1/3$ that is
decoupled from neutral modes: 
\begin{equation}
\chi_{111}=\frac{\phi_{{\rm c}}}{\sqrt{3}}
\end{equation}
also there are three ``neutralons'' with $e^{*}=0$ and $\Delta=2/3$:
\begin{equation}
\chi_{110},\chi_{10-1},\chi_{011}
\end{equation}
and conjugate. Interestingly there are six quasi-particles with $e^{*}=1/3$
and $\Delta=1$ (free fermion scaling dimension): 
\begin{equation}
\chi_{001},\chi_{100},\chi_{012},\chi_{210},\chi_{122},\chi_{221}\label{good}
\end{equation}
These could be a signature of $\nu=1$ like condensate (plasmon waves
on the boundary of inner condensate of $e^{*}=1/3$ quasi-particles).
These last quasi-particles (\ref{good}) have non-zero charge and
are coupled to the upstream modes. Therefore they are responsible
for the dominant contribution to the thermo-electric effects upstream.

\begin{figure}[h]
\centering{}\includegraphics[width=0.6\textwidth]{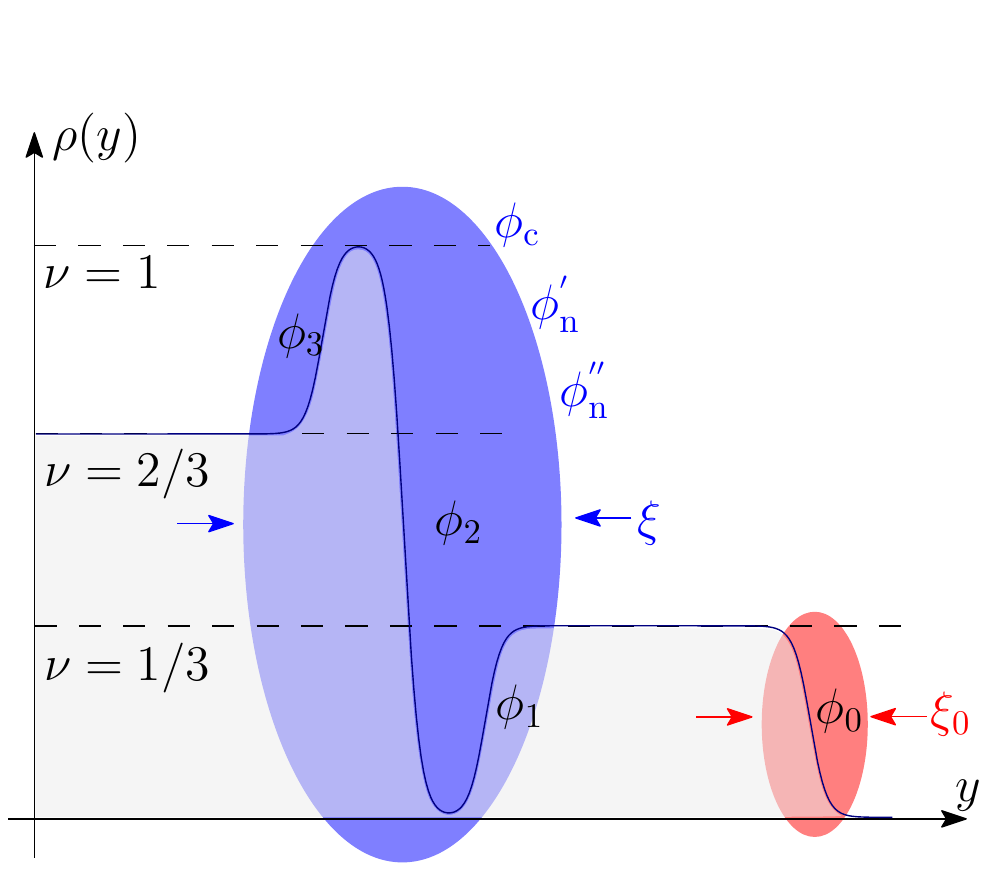} \caption{\textbf{Theoretical edge state structure:} Schematic of the effective
edge model. The blue curve shows possible transversal profile of charge
density near the boundary of two-dimensional electron gas. In the
effective theory, the inner modes are described in a universal way
by three boson fields $\phi_{{\rm c}},\phi'_{{\rm n}},\phi''_{{\rm n}}$
with action (\ref{action}). Note that these modes have larger width
than the outermost charged mode.}
\label{scheme-fig} 
\end{figure}

\subsection*{Quantitative results: Thermo-electric current}

Upstream thermo-electric current through a single level with energy
$\epsilon_{0}$ can be estimated as 
\begin{equation}
I_{{\rm th-el}}\propto[f_{D}(\epsilon_{0})f_{U}(-\epsilon_{0})-f_{U}(\epsilon_{0})f_{D}(-\epsilon_{0})]\label{cur-def}
\end{equation}
Here effective occupation numbers $f_{i}(\epsilon_{0})$ are simply
Fourier transforms of the corresponding correlation functions: \begin{subequations}
\begin{eqnarray}
f(\epsilon_{0}) & = & \int dte^{i\epsilon_{0}t}K(t)\\
K(t) & = & \langle\exp[-i\chi_{\mathbf{n}}(t)]\exp[i\chi_{\mathbf{n}}(0)]\rangle
\end{eqnarray}
\label{upper} \end{subequations}

We take into account that downstream modes on the upper edge of quantum
dot originate from a cold reservoir with base temperature $T_{0}$,
while upstream modes are from \char`\"{}hot\char`\"{} Ohmic contact
with temperature $T_{1}$. Assuming that the dominant cooling mechanism
is by four edge states \citep{Takei2011,Slobodeniuk}, and taking
into account that every chiral bosonic mode at temperature $T$ carries
a heat flux $\pi T^{2}/12$ we could write down the heat balance equation
between the heat produced in the contact and heat dissipated (carried
away by edge modes): 
\begin{equation}
I\Delta\mu=\frac{\pi}{3}\left(T_{1}^{2}-T_{0}^{2}\right).
\end{equation}
Therefore we find that $T_{1}=\sqrt{T_{0}^{2}+(\Delta\mu/\pi)^{2}}$
for $\nu=2/3$ and we have: 
\begin{equation}
f_{U}(\epsilon_{0})=\int dte^{i\epsilon_{0}t}\left[\frac{T_{0}}{\sinh(\pi T_{0}(t-i\eta))}\right]^{\delta_{0}}\left[\frac{T_{1}}{\sinh(\pi T_{1}(t-i\eta))}\right]^{\delta_{1}}.
\end{equation}
In our model $\delta_{0}=1/3,\delta_{1}=2/3$ and $\Delta=\delta_{0}+\delta_{1}=1$.
For the lower edge effective occupation coincides with free fermions,
since $\Delta=1$: 
\begin{equation}
f_{D}(\epsilon_{0})=\int dte^{i\epsilon_{0}t}\frac{T_{0}}{\sinh(\pi T_{0}(t-i\eta))}=\frac{1}{e^{\epsilon_{0}/T}+1}
\end{equation}
We could also estimate the upper effective occupation (\ref{upper})
taking into account that at small energies the integral comes from
the large times, where the correlation function decays exponentially:
\begin{equation}
K(t)\simeq\exp[-\pi T_{{\rm eff}}|t|],\hspace{12pt}T_{{\rm eff}}=\frac{T_{0}+2T_{1}}{3}.
\end{equation}
Thus one could make an estimation: 
\begin{equation}
f_{U}(\epsilon_{0})\sim\frac{1}{e^{\epsilon_{0}/T_{{\rm eff}}}+1}
\end{equation}
Results of this approximation as compared to exact numeric integration
are shown in Fig 2d in the main text . They also agree well with the
experimental data.

The asymptotic behavior of the thermo-electric current (\ref{cur-def})
in effective temperature approximation is \begin{subequations} 
\begin{eqnarray}
I_{{\rm th-el}} & \sim & \Delta\mu^{2}/T_{0}^{2},\hspace{12pt}\Delta\mu\ll T_{0}\\
I_{{\rm th-el}} & \sim & {\rm const},\hspace{12pt}\Delta\mu\gg T_{0}
\end{eqnarray}
\end{subequations} The saturation constant itself does not depend
on base temperature $T_{0}$ when it is sufficiently low, but is suppressed
as $\delta\epsilon_{0}/T_{0}$ when temperature becomes comparable
with the level spacing $\delta\epsilon_{0}$ of the quantum dot.

\subsection*{Qualitative discussion: QPC charge and neutral transmissions}

The key ingredient for the results of previous section is the electron-like
scaling behavior of fractional quasi-particles (\ref{good}). Here
we also use the argument that the fractional quasi-particles coupled
to neutral modes behave like free electrons. It is important to note
that the effective action (\ref{action}) is an intermediate fixed
point, it has relevant neutralons $\Delta=2/3$. In contrast, low-energy
fixed point of Polchinski-Kane-Fisher model \citep{Kane1994} has
only irrelevant neutralons with $\Delta=2$. However, quantum point
contacts and quantum dots can give additional energy scale that pins
the intermediate fixed point.

There are three wide inner states in our model: one charged and two
upstream neutral and one narrow charged outer. The transmission of
a single channel could be modeled in the quasi-classical approximation
\citep{Landau1977,Buttiker1990} as 
\begin{equation}
t_{i}(V)=\frac{1}{1+e^{(V_{i}-V)/\delta V_{i}}}
\end{equation}
where $i={\rm in,out}$. In our model there are two channels that
are spatially separated, therefore they will be pinched off at different
QPC voltages: 
\begin{equation}
t_{{\rm charge}}(V)=\frac{1}{2}\left[\frac{1}{1+e^{(V_{{\rm in}}-V)/\delta V_{{\rm in}}}}+\frac{1}{1+e^{(V_{{\rm out}}-V)/\delta V_{{\rm out}}}}\right]\label{t-ch}
\end{equation}
Analogously, both neutral modes are located at inner channel, so that:
\begin{equation}
t_{{\rm neutral}}(V)=\frac{1}{1+e^{(V_{{\rm in}}-V)/\delta V_{{\rm in}}}}\label{t_neutr}
\end{equation}
It is natural to assume that $\delta V_{{\rm in}}/\delta V_{{\rm out}}\propto\xi/\xi_{0}\gg1$.
\begin{itemize}
\item Appearance of quasi-particles with charge $e^{*}=1/3$ but with electronic
scaling dimension $\Delta=1$. 
\item Couplings of the above quasi-particles to the changed and neutral
modes are universal $\delta_{0}=1/3$ and $\delta_{1}=2/3$ in the
limit of strong Coulomb interactions. 
\item Inner and outer edge channels have significantly different widths
$\xi/\xi_{0}\gg1$. 
\item Upstream neutral excitation only appear in the inner channel. 
\end{itemize}
Most of these features will appear in all other filling fractions
that allow hierarchic condensates. And indeed, the experimental data
for $\nu=3/5$ show that all the results are very similar.

\renewcommand\refname{Supplementary References}\bibliographystyle{unsrt}